\documentstyle[graphicx,natbib,epsfig,amsmath,amssymb,traditabstract,letter]{aa}

\newcommand{\tex}{\,\hbox{$T_{\rm ex}$}}

\newcommand{\msun}{\,\hbox{$M_{\odot}$}}
\newcommand{\lsun}{\,\hbox{$L_{\odot}$}}
\newcommand{\kms}{\,\hbox{\hbox{km}\,\hbox{s}$^{-1}$}}

\newcommand{\htwo}{\,\hbox{$\rm{H_ 2}$}}

\newcommand{\feii}{\,\hbox{[\ion{Fe}{II}}]}

\newcommand{\fft}{\hbox{$\rm F_{43}/F_{21}$}}

\newcommand{\coto}{\hbox{$\rm CO(2$-$1)$}}

\newcommand{\coft}{\hbox{$\rm CO(4$-$3)$}}

\begin{document}
\textheight=24.8 cm
\addtolength{\topmargin}{-.2cm}
\setlength{\parskip}{0.0mm plus 0.0mm minus0.0mm}

   \title{  ALMA reveals optically thin, highly excited CO gas in the jet-driven winds of the galaxy IC5063 }
 
   \subtitle{}

   \author{K. M. Dasyra\inst{1}, 
	 F.  Combes\inst{2,3},
	 T. Oosterloo\inst{4,5},
	 J. B. R. Oonk\inst{4,6},
	 R. Morganti\inst{4,5},
	 P. Salom\'e\inst{2},
	 \and
	 N. Vlahakis\inst{1}
	  }

   \institute{
		Department of Astrophysics, Astronomy \& Mechanics, Faculty of Physics, National and Kapodistrian University of Athens, Panepistimiopolis Zografou, 15784, Greece
		\and	
		LERMA, Observatoire de Paris, CNRS, UPMC, PSL Univ., Sorbonne Univ., 75014 Paris, France
		\and
		Coll\`ege de France, 11 place Marcelin Berthelot, 75005, Paris, France
		\and	
		ASTRON, the Netherlands Institute for Radio Astronomy, Postbus 2, 7990 AA, Dwingeloo, The Netherlands 
		\and
		Kapteyn Astronomical Institute, University of Groningen, Postbus 800, 9700 AV Groningen, The Netherlands
		\and
		Leiden Observatory, Leiden University, PO Box 9513, 2300 RA Leiden, The Netherlands
             }

   \date{}

    \abstract 
    { Using CO (4-3) and (2-1) Atacama Large Millimeter Array (ALMA) data, we prove that the molecular gas in the jet-driven winds of the galaxy IC5063 is more highly excited than the rest of the molecular gas in the disk of the same galaxy.  On average, the \coft /\coto\ flux ratio is 1 for the disk and 5 for the jet accelerated or impacted gas. Spatially-resolved maps reveal that in regions associated with winds, the \coft /\coto\ flux ratio significantly exceeds the upper limit of 4 for optically thick gas. It frequently takes values between 5 and 11, and it occasionally further approaches the upper limit of 16 for optically thin gas. Excitation temperatures of 30-100\,K are common for the molecules in these regions. If all of the outflowing molecular gas is optically thin, at 30-50\,K, then its mass is 2$\times$10$^6$\msun . This lower mass limit is an order of magnitude below the mass derived from the \coto\ flux in the case of optically thick gas. Molecular winds can thus be less massive, but more easily detectable at high z than they were previously thought to be.
 }


   \titlerunning{ Optically thin, highly excited CO in the jet-driven winds of IC5063  }
   \authorrunning{Dasyra et al.}
   
\maketitle

\section{Introduction}
\label{sec:intro}

Jets launched by supermassive black holes are presently being evaluated as important mechanisms for galaxy evolution, affecting multiple phases of the interstellar medium (ISM). By depositing energy in the hot and tenuous gas, jets are proposed to delay the inflow of intergalactic gas into ellipticals and moderate their growth \citep{bower06, croton06}. By depositing energy in the cold and dense gas, jets are proposed to influence the star-formation properties of disks \citep{wagner11, gaibler12}. Observational evidence that the impact of a jet upon dense clouds initiates molecular gas winds has been successfully gathered \citep[e.g.,][]{morganti05, rupke11, sakamoto14, garcia-burillo14, dasyra15, aalto16}. 

In this letter, we provide observational evidence that the molecular gas accelerated or impacted by a jet is more highly excited and less optically thick than the rest of the ISM. We prove this for the galaxy IC5063, an elliptical with a molecular disk in its center. The jet launched from the black hole of IC5063 propagates through the disk, nearly parallel to its plane. Once the jet encounters clouds, it scatters gas, forming atomic and molecular winds \citep{oosterloo00,morganti07,morganti13,tadhunter14}. Using ultra-deep ESO VLT near infrared (NIR) data, we identified at least four discrete starting points of ionized and warm molecular gas winds near the jet trail \citep{dasyra15}. This constitutes the highest number of known jet-driven winds in a galaxy. The winds sweep the galaxy's inner kpc$^2$, allowing for a spatially-resolved study of the accelerated gas properties. Atacama Large Millimeter Array (ALMA) data have revealed the presence of cold molecular gas in most of these winds \citep{morganti15}.

\section{The ALMA data}
\label{sec:data}
New, band 8 data of IC5063 were taken with ALMA on May 18 2016 for the program 2015.1.00420.S (PI Combes). Forty two antennas observed the galaxy for 39 minutes on-source. Three 1.875\,GHz-bandwidth spectral windows were employed. Two of the spectral windows were partly overlapping to cover the $\pm$1000\kms\ range around \coft . The third spectral window was used for duplication purposes in the $\pm$300\kms\ range: its role was to ensure that no significant continuum offset exists between the former windows due to flux and bandpass calibration uncertainties. The data were reduced within the Common Astronomy Software Applications (CASA) environment, version 4.6.0. The calibration script was executed as delivered by ESO. Pallas, J1924-2914, and J2056-4714 were respectively used as flux, bandpass, and phase calibrators. The flux calibration accuracy was better than 5\%, as cross-checked with J1924-2914, which is not variable as J2056-4714 is. To construct images from visibilities, we chose a 25\kms\ spectral bin and a 0.06\arcsec\ pixel scale. The synthesized beam size was 0.4\arcsec $\times$0.4\arcsec . 

To study the gas excitation, we also used \coto\ data from the program 2012.1.00435.S \citep{morganti15}. Thirty one antennas observed IC5063 in band 6 for 30 mins on-source. J2056-4714 was used as bandpass and phase calibrator, Neptune as flux calibrator. Four 1.875\,GHz-bandwidth spectral windows were employed, with one of them covering the $\pm$1000\kms\ range around \coto. The flux calibration was performed in CASA 4.2.2. Imaging was carried out in CASA 4.6.0 (because of continuum subtraction improvements). The cube was created at a 25\kms\ spectral bin and at a 0.1\arcsec\ pixel scale. The synthesized beam was 0.56\arcsec $\times$0.53\arcsec . 

All images were reconstructed using Briggs robust=0.5 weighting, and primary-beam corrected. No important flux losses were considered for them: the {\it uv} plane coverage was high and the area under examination  was small (3-5\arcsec\ in individual channels) compared to the primary beam in both bands (28\arcsec,\,14\arcsec ). To compare them, we re-gridded the \coto\ cube to the pixel scale of the \coft\ cube, and we smoothed by Gaussian convolution the \coft\ cube to the resolution of the \coto\ beam. In the common resolution, the noise is 0.26 mJy/beam for \coto , and 3.4 mJy/beam for \coft .

\section{Results: gas excitation differences in the wind and in the ambient medium}
\label{sec:results}

\begin{figure*}[h!]
\begin{center}
\includegraphics[width=19cm]{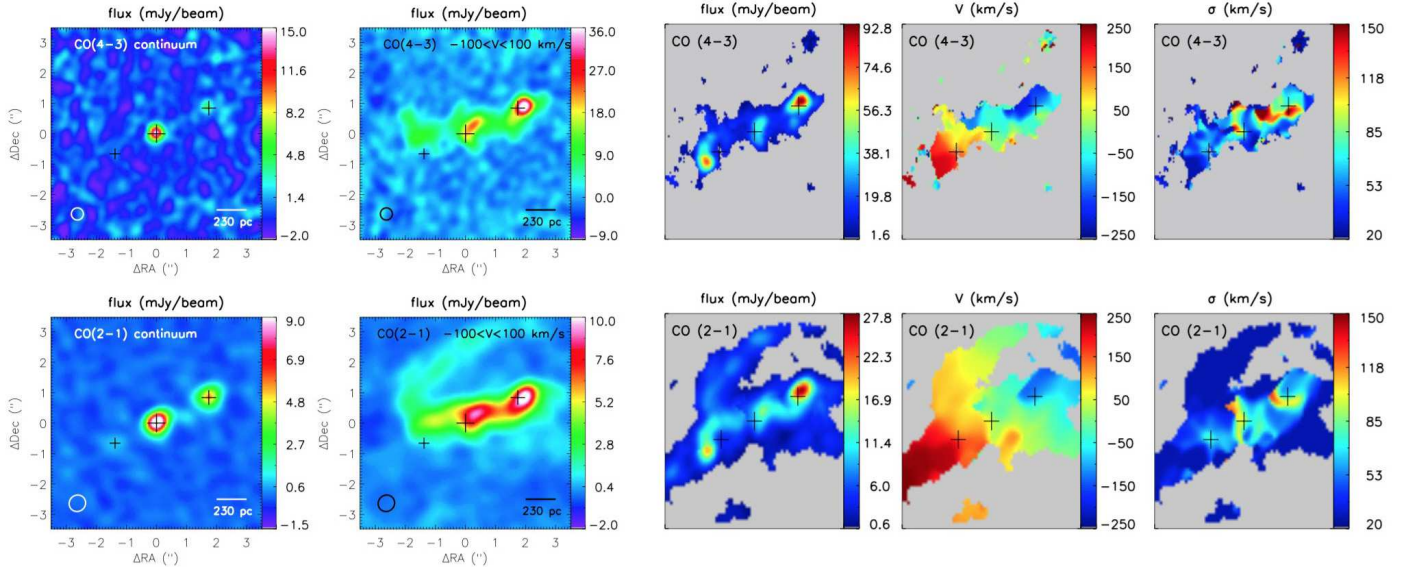} \\
\caption{ {\it Left:} images of CO (4-3), (2-1), and their nearby continua, extracted from the optimal-resolution cubes (i.e., 0.4\arcsec$\times$0.4\arcsec\ beam in band 8 and 0.56\arcsec$\times$0.53\arcsec\ beam in band 6). The line fluxes in mJy/beam were averaged over the -100$<$V$<$100\kms\ range. Crosses indicate the locations of the nucleus and the two radio lobes, as seen in the \coto\ continuum (see Sect.~\ref{sec:data} for the coordinates). {\it Right:} maps of the amplitude, velocity, and velocity dispersion of a Gaussian line-profile fit. All panels are centered at the nucleus, at 20:52:02.37 -57:04:07.6. They have a field of view (FOV) of 7\arcsec$\times$7\arcsec .}
\label{fig:basics}
\end{center}
\end{figure*}

Images of the spatially-resolved line emission are shown in Fig.~\ref{fig:basics}, following the subtraction of the (also spatially-resolved) continuum. \coto\ is bright along the jet propagation axis and in spirals arms, forming the S-shaped feature identified by \citet{morganti15}. Contrarily,  \coft\ is mainly detected along the jet axis. In both lines, the maximum emission is detected near the north-west radio lobe, the south-east radio lobe, and the radio core. Gaussian modeling of the emission in both lines indicates that the projected rotational velocity reaches $\sim$200\kms\ within 600\,pc from the nucleus. High  \coft\  velocity dispersion (either real or caused by the projection of several kinematic components) is seen north of the nucleus, near the north-west radio lobe, and along the biconical outflow detected by \citet{dasyra15} in NIR \htwo\ and \feii\ data. The latter outflow's starting point is mid-way between the nucleus and the north-west lobe, and its axis is oriented perpendicularly to the jet trail.

The total flux detected in \coft\ is 118.0\,$\pm$\,5.0\, Jy\kms , whereas\, that detected\, in \coto\ is 70.2 $\pm$ 4.7 Jy\kms .  The \coto\ flux is 30\% lower than that measured from single-dish, APEX data. The difference is attributed to uncertainties in the continuum determination in the APEX data, in which the blue CO wing was covering most of the available band \citep{morganti13, morganti15}.  

\begin{figure*}[h!]
\begin{center}
\includegraphics[width=0.78\columnwidth]{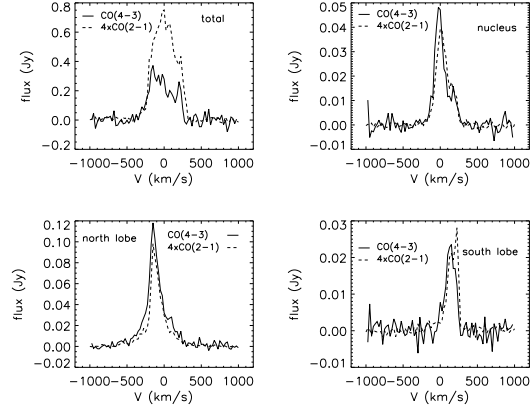} 
\caption{ Integrated \coft\ and \coto\ spectra, extracted from the cubes at the common spatial resolution. The \coto\ flux is multiplied by a factor of four to facilitate comparison with the \coto\ flux.  {\it Upper left:} Emission integrated over the whole disk, that is within an ellipse with a 12\arcsec\ major axis, 4\arcsec\ minor axis, and \hbox{-65$^{\rm o}$} position angle.  {\it Upper right and bottom:} Emission integrated within a beam (i.e., within a radius of 0.28\arcsec ) from the nucleus, and from the north and south radio lobes. }
\label{fig:spectra}
\end{center}
\end{figure*}

The overall  \coft /\coto\ flux ratio, \fft , is indicative of a sub-thermally excited gas reservoir (see Section~\ref{sec:discussion}). However, when integrated within a beam's radius from the nucleus (at 20:52:02.37 -57:04:07.6) and the radio lobes (at 20:52:02.15 -57:04:06.7 and  20:52:02.54 -57:04:08.3), the flux in \coft\  is $\sim$4 times higher than that in \coto\ (Fig.~\ref{fig:spectra}).

To derive the fraction of the flux originating from the jet accelerated or impacted gas, we modeled the disk emission and subtracted it from the observed spectrum in each pixel. A Gaussian fitting algorithm was employed, prefered over a two-component Gaussian fitting algorithm for its lower errors. The disk velocity $V$ was tracing the peak of the line emission, never exceeding $\pm$250\kms\ in the inner 1\,kpc$^2$. The velocity dispersion $\sigma$ was constrained to values below 50\kms : this is the average $\sigma$ for both lines in regions that are located more than a beam away from the nucleus and the radio lobes. In the residual cubes, the integrated flux is 12.3($\pm$2.5) Jy\kms\ for \coto , and 59.7($\pm$9.2) Jy\kms\ for \coft . This result indicates that $\sim$half of the reservoir probed by \coft\ is kinematically or excitationally affected by the jet. The \fft\ ratio is 4.9($\pm$)1.7 for this gas,\,and 1.0($\pm$)0.4 for the rest of the gas in the disk. For the fastest gas in the wind, which is moving with -600$<$V$<$-400\kms\ and which is seen on the north-west radio lobe, \fft\ is 8.0($\pm$2.4).

\begin{figure*}[h!]
\begin{center}
\includegraphics[width=0.9\textwidth]{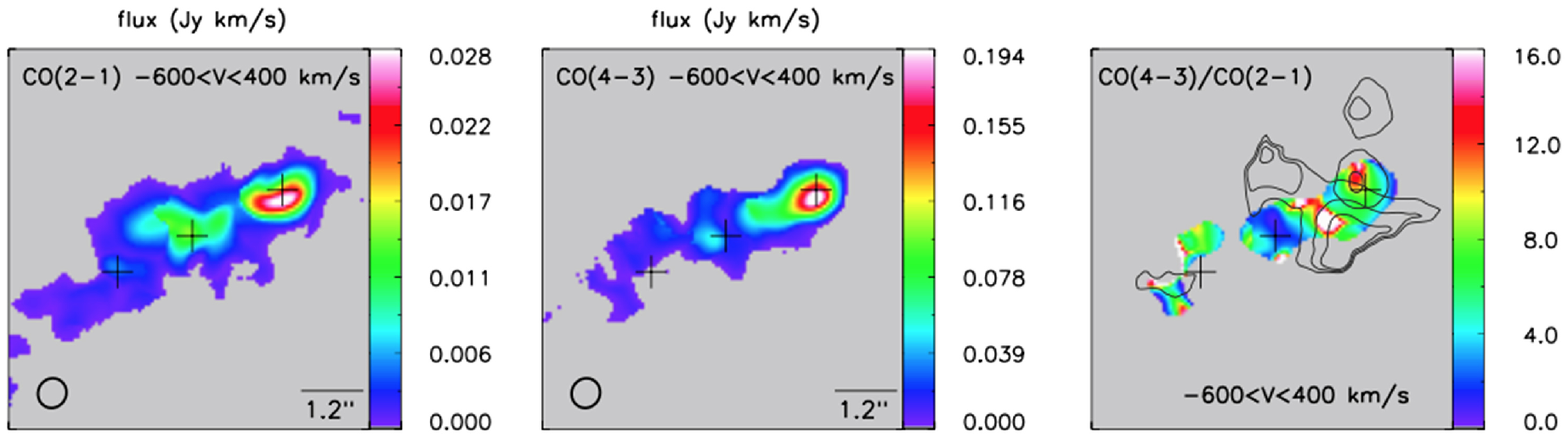} \\ 
\caption{ {\it Left, middle:} residual images from the subtraction of a disk model from the data (see Section~\ref{sec:results}). Their fluxes, integrated in the -600$<$V$<$400\kms\ range, are in units of Jy\kms\ per pixel. The common beam is 0.56\arcsec$\times$0.53\arcsec . Pixels with signal-to-noise (S/N)\,$<$5 have been masked. {\it Right:} \fft\ of the residual images. Contours of the outflowing-to-ambient \feii\ emission from \citet{dasyra15} are overplotted, in steps of 0.2 starting from 0.1 for the fast winds near the lobes, and steps of 0.33 starting from 0.33 for the intermediate-velocity (component) of the other winds. The FOV and center are as in Fig.~\ref{fig:basics}.}
\label{fig:residual_maps}
\end{center}
\end{figure*}

\begin{figure*}[h!]
\begin{center}
\includegraphics[width=0.9\textwidth]{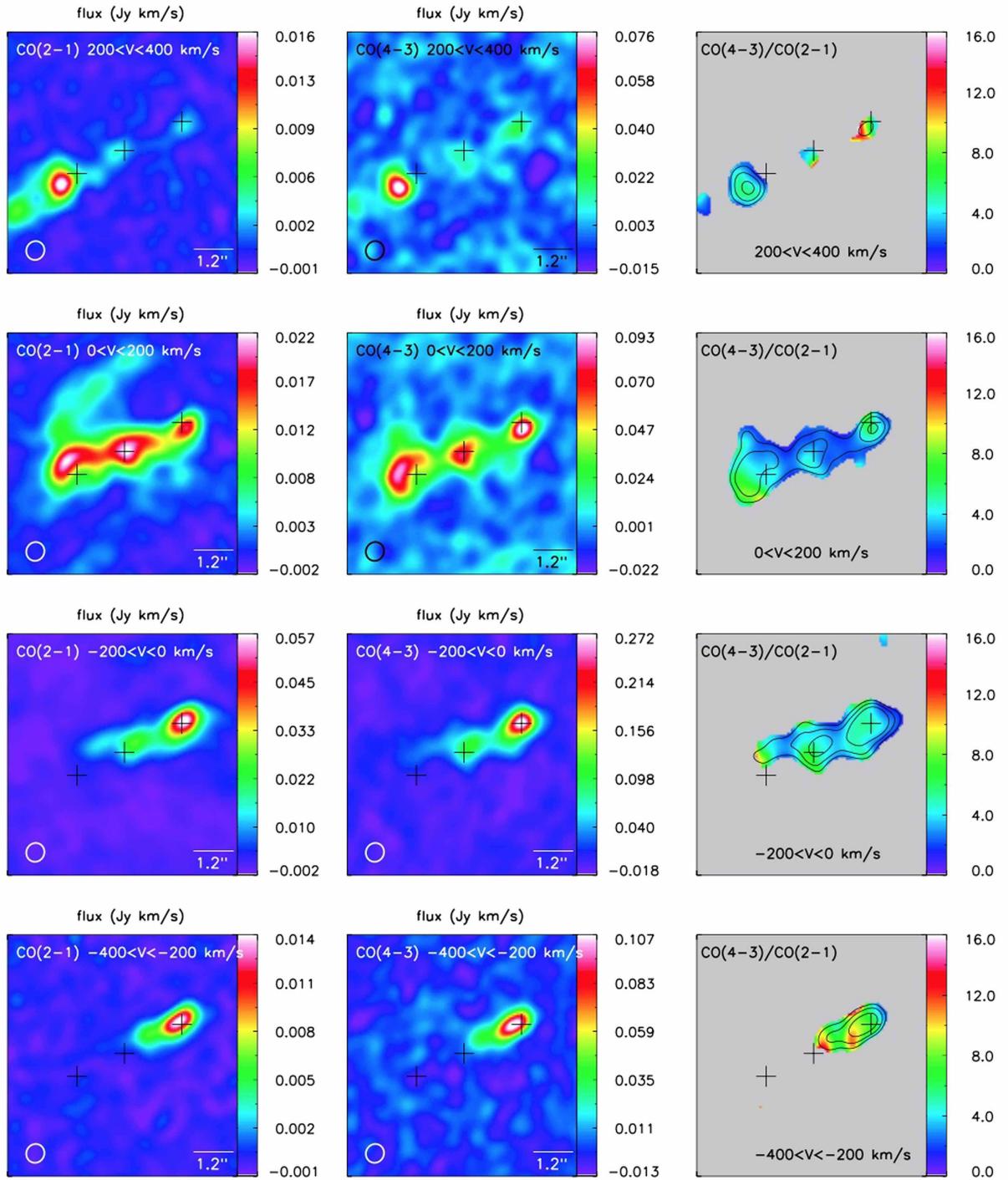} \\ 
\caption{ CO (4-3) and (2-1) images and \fft\ ratio, integrated in 200\kms\ bins. All panels were constructed directly from the data, and, unlike Fig.~\ref{fig:residual_maps}, they contain the disk emission. The image units, beam, FOV, and center are as in Fig.~\ref{fig:residual_maps}. Pixels with S/N$<$3 have been masked. The contours in the last panel indicate where the \coft\ emission reaches S/N of 5,\,10,\,and 20.} 
\label{fig:ratio_maps}
\end{center}
\end{figure*}

Regions with high CO excitation are shown in a \fft\ map that we constructed after collapsing the disk-subtracted cubes in the \hbox{-600}$<$V$<$400\kms\ range (Fig. ~\ref{fig:residual_maps}). These regions often coincide with wind starting points known from NIR \htwo\ and \feii\ data \citep{tadhunter14,dasyra15}. Values $\ge$8 are seen north of the north-west radio lobe, where the \htwo\ is also highly excited \citep{dasyra15}, and where diffuse 17.8 GHz emission is detected \citep{morganti07}. Values approaching 16 are seen in blobs that nearly coincide with the biconical outflow axis perpendicular to the jet \citep[see the line defined by R1 and R2 in Fig. 8 of ][]{dasyra15}. Within 200\,pc from the nucleus, the wind associated with \fft$\ge$6 regions was seen in \coto\ by \citet{morganti15}. East of the south-east radio lobe, high-velocity Fe ions and highly excited CO molecules co-exist (Fig. ~\ref{fig:residual_maps}). Counter-rotating gas is detected near the \fft $\sim$7 region north of the same radio lobe in both CO lines (Fig.~\ref{fig:ratio_maps}). 
All the above-mentioned regions are also seen in \fft\ maps that we created from the cubes prior to the disk subtraction (Fig.~\ref{fig:ratio_maps}).

\section{Discussion: gas mass in the wind}
\label{sec:discussion}

Our analysis indicates the presence of at least two distinct molecular gas components: a low-excitation component associated with clouds in the disk, and a high-excitation component associated with gas in the jet-driven winds or near jet-cloud impact points. To be bright in \coft, which has a critical density of 1.7$\times$10$^5$ cm$^{-3}$, both components must contain clouds of high volume density. However, while the low excitation component is optically thick, the high excitation component is (partly) optically thin: \fft\ values $>$4 can only be consistent with emission from optically thin gas at high temperatures. 

From the antenna temperature definition, assuming that the molecular gas is in local thermodynamic equilibrium (LTE) and that its excitation temperature \tex\ greatly exceeds that of the cosmic microwave background (see below), we obtain analytic solutions for \fft\ (with fluxes in units of Jy\kms ). For optically thick gas, \fft\ = ${ \rm (\nu_{43}/\nu_{21})^3 [e^{(h \nu_{21} /kT_{ex})}-1]/[e^{(h \nu_{43} /kT_{ex})}-1]}$. An upper limit of 4 is reached at the Rayleigh-Jeans limit (as expected from the same line brightness temperature). For optically thin gas, \fft\ = ${\rm (\nu_{43}/\nu_{21})^4 e^{-7h \nu_{10} /kT_{ex}}}$. An asymptotic value of 16 is obtained for  \tex $>$$>$7h$\nu_{10}$/k $\sim$ 39 K. From this relation, we find that \tex =32K, if all of the gas kinematically or excitationally affected by the jet (with \fft =5) is optically thin. For the fastest gas in the wind (with \fft =8),
\tex\ is 56K. Near the biconical outflow base, where \fft\ is 11, \tex\ is 100K. The temperatures of the gas in the wind can thus be quite high for LTE, and they can be even higher for non-LTE.
 
It is unlikely that this result reflects infrared (IR) pumping of the populations of quantum states with high rotational number J. The molecules should see an IR source of $>$160K with a large filling factor \citep{carroll81} in all pertinent regions of Figs.~\ref{fig:residual_maps} and \ref{fig:ratio_maps}. Infrared pumping has been deemed unimportant for HCN in the wind of Mrk231 \citep{aalto12}. The high \tex\ is instead related to the creation of dense fragments once the jet impinges on molecular clouds. The contribution of the jet to the dissipation and heating of the clouds could thus have a negative impact on the overall star formation of IC5063. Still, it doesn't rule out a local enhancement of star formation in compressed clouds. 

Given the above bimodality, we determined the gas mass in the wind and in the disk using the appropriate CO-luminosity-to-\htwo -mass conversion factor, $\alpha_{CO}$, for optically thin and thick emission, respectively. For the optically thick gas, we use Eq. 3 of \citet{solomon97} with $\alpha_{CO}$=4.6\msun /(K\kms\,pc$^2$), as the far infrared luminosity of IC5063 is 2$\times$10$^{10}$\lsun .  For \hbox{$\rm F_{21}/F_{10}$}=4, a mass of 4.0($\pm$0.5)$\times$10$^8$\msun\ is obtained for the disk. For the optically thin gas, \hbox{$\rm F_{21}/F_{10}$} = $ {\rm (\nu_{21}/\nu_{10})^4 e^{-2h \nu_{10} /kT_{ex}}}$ and $\alpha_{CO}=0.26\,(\tex /30)\,e^{5.53/\tex-0.184}$ \msun / (K\kms\,pc$^2$), for a Galactic CO/\htwo\ abundance \citep{bolatto13}. \hbox{For $\rm F_{21}$=12.3($\pm$2.5) Jy\kms\ and \fft=5, then $\alpha_{CO}$} \hbox{is 0.27\msun/\,(K\kms\,pc$^2$) and the wind mass  is 1.7($\pm$0.3)}  \hbox{$\times$10$^6$\msun . For \fft=8, $\alpha_{CO}$ is 0.44\msun /\,(K\kms\,pc$^2$)} and the wind mass  is 2.4($\pm$0.5)$\times$10$^6$\msun . An order-of-magnitude difference in $\alpha_{CO}$ between the disk and the wind was already assumed by \citet{morganti15}, but the use of an optically-thick \hbox{$\rm F_{21}/F_{10}$} value led to a wind mass of 1.9$\times$10$^7$\msun . Our current knowledge that the wind is partly optically thin leads us to revise its molecular content down: 2$\times$10$^6$\msun\ is its lower mass limit, if all of its gas is optically thin. This is comparable to the atomic wind mass \citep[3.6$\times$10$^6$\msun ; ][]{morganti07}. The actual molecular wind mass will be derived when all data, including new CO (1-0) and (3-2) observations, are analyzed assuming non-LTE conditions: when both high-temperature gas and low-temperature gas (contributing to low J lines) are used to fit the CO spectral line energy distribution (SLED). Still, our simplified LTE analysis proves that molecular wind masses can be overestimated \citep[contrary to what is known from Mrk231;][]{cicone12}.

\section{Summary}
\label{sec:conclusions}

We obtained new ALMA \coft\ data and combined them with previous \coto\ data to study the molecular gas excitation in the jet-driven winds of IC5063. We found that \coft\ is mainly detected along the jet. Half of its emission comes from gas that was impacted by the jet, while only one sixth of the \coto\ emission comes from the same gas. In this gas, the average \fft\ ratio is 5. In regions with a strong jet-cloud interaction, the ratio reaches 12 in the original cubes and approaches 16 in the disk-subtracted cubes. Contrarily, the same ratio for the rest of the gas in the disk is 1. We thus detect at least two gas components: one optically thick related to the disk, and another optically thin related to the winds. Because the outflowing molecular gas is partly optically thin, its mass is lower than that previously thought, as low as 2$\times$10$^6$\msun . This is comparable to the mass of the outflowing HI.  Excitation temperatures of 30-100K are common for the outflowing molecules. The skewing of the CO SLED toward higher J numbers can facilitate the detection of winds at high z.

\begin{acknowledgements}
K. D. acknowledges support by the European Commission through a Marie Curie Intra-European Fellowship (PIEF-GA-2013-627195;  'BHs shaping galaxies') awarded under the 7th Framework Programme. R. M. gratefully acknowledges support from the European Research Council under the European Union's 7th Framework Programme, ERC Advanced Grant RADIOLIFE-320745.
This paper makes use of the ALMA data 2015.1.00420.S and 2012.1.00435.S.  ALMA is a partnership of ESO (representing its member states), NSF (USA), NINS (Japan), together with NRC (Canada), NSC and 
ASIAA (Taiwan) and KASI (Republic of Korea), in cooperation with the Republic of Chile.  The Joint ALMA Observatory is operated by ESO, AUI/NRAO and NAOJ.
\end{acknowledgements}

{}


\begin{thebibliography}{}

\bibitem[Aalto et al.(2012)]{aalto12}
Aalto, S., Garcia-Burillo, S., Muller, S. et al.  2012, A\&A, 537, 44

\bibitem[Aalto et al.(2016)]{aalto16}
Aalto, S., Costagliola, F., Muller, S., et al. 2015, A\&A, 590, 73

\bibitem[Bolatto et al.(2013)]{bolatto13}
Bolatto, A., Wolfire, M., \& Leroy, A., 2013 ARA\&A, 51, 207

\bibitem[Bower et al.(2006)]{bower06}
Bower, R. G., Benson, A., Malbon, R., et al. 2006, MNRAS, 370, 645

\bibitem[Carroll \& Goldsmith(1981)]{carroll81}
Carroll, T. J., \& Goldsmith, P. F. 1981, ApJ, 245, 891

\bibitem[Cicone et al.(2012)]{cicone12}
Cicone, C., Feruglio, C., Maiolino, R., et al. 2012, A\&A, 543, 99

\bibitem[Croton et al.(2006)]{croton06} 
Croton, D. J., Springel, V., White, S., et al. 2006, MNRAS, 367, 864 

\bibitem[Dasyra \& Combes(2012)]{dasyra_combes12}
Dasyra, K. M., \& Combes, F. 2012, A\&A, 541, L7

\bibitem[Dasyra et al.(2014)]{dasyra14}
Dasyra, K. M., Combes, F., Novak, G., et al. 2014, A\&A, 565, 46

\bibitem[Dasyra et al.(2015)]{dasyra15}
Dasyra, K. M., Bostrom, A., Combes, F., et al. 2015, ApJ, 815, 34

\bibitem[Gaibler et al.(2012)]{gaibler12} 	
Gaibler, V., Khochfar, S., Krause, M., et al. 2012, MNRAS, 425, 438

\bibitem[Garc\'ia-Burillo et al.(2014)]{garcia-burillo14}
Garc\'ia-Burillo, S., Combes, F., Usero, A., et al. 2014 A\&A, 567, 125

\bibitem[Morganti et al.(2005)]{morganti05} 
Morganti, R., Tadhunter, C., \& Oosterloo, T.  2005, A\&A, 444, L9

\bibitem[Morganti et al.(2007)]{morganti07} 
Morganti, R., Holt, J., Saripalli, L., et al. 2007,\,A\&A, 476, 735

\bibitem[Morganti et al.(2013)]{morganti13} 
Morganti, R., Frieswijk, W., Oonk, R., et al.  2013, A\&A, 552, L4	

\bibitem[Morganti et al.(2015)]{morganti15}
Morganti, R., Oosterloo, O, Oonk, R., et al., 2015 A\&A, 580, 1

\bibitem[Oosterloo et al.(2000)]{oosterloo00}
Oosterloo, T., Morganti, R., Tzioumis, A., et al. 2000, AJ, 119, 2085

\bibitem[Rupke \& Veilleux(2011)]{rupke11}
Rupke, D. S. N., \& Veilleux, S. 2011, ApJ, 729, L27

\bibitem[Sakamoto et al.(2014)]{sakamoto14}
Sakamoto, K. Aalto, S., Combes, F., et al. 2014, ApJ, 797, 90  

\bibitem[Solomon et al.(1997)]{solomon97}
Solomon, P., Downes, D., Radford, S., et al. 1997, ApJ, 478, 144

\bibitem[Tadhunter et al.(2014)]{tadhunter14}
Tadhunter, C., Morganti, R., Rose, M., et al. 2014, Nature, 511, 440

\bibitem[Wagner \& Bicknell(2011)]{wagner11}
Wagner, A. Y., \& Bicknell, G. V. 2011, ApJ, 728, 29


\end{thebibliography}
\end{document}